# Knowledge Representation for High-Level Norms and Violation Inference in Logic Programming

B.O. AKINKUNMI[1] and Moyin F. BABALOLA[2]

[1]Dept of Computer Science,

University of Ibadan, Ibadan, Nigeria.

[2]Dept of Computer Studies,

The Polytechnic, Ibadan, Nigeria.

**ABSTRACT**

Most of the Knowledge Representation formalisms developed for representing prescriptive norms in multi-agent normative systems can be categorized as either suitable for representing high-level or low level norms. We argue that low-level norm representation do not advance the cause of autonomy in agents in the sense that it is not the agent itself determining the normative position it should be in at a particular time, on the account of a more general rule. In other words the agent depends on some external system for a *nitty-gritty* prescription of its obligations and prohibitions. On the other hand, high-level norms which have an explicit description of a norm's precondition and have some form of implication, do not, as they exist in the literature, support generalized inferences about violation like low level norm representation formalisms do.

This paper presents a logical formalism for representation of high level norms in open societies that enable violation inferences that detail the situation in which the norm violation took place and the identity of the norm violated.

Norms are formalized as logic programs, whose heads each specify what an agent is obliged or permitted to do when some situation arises and within what time constraint of the situation. Each norm is also assigned an identity, using a reification scheme. The body of each logic programs usually describes the nature of the situation in which the agent is expected to act or desist from acting. This kind of violation inference is novel in the literature.

## 1. INTRODUCTION

Norms are used in MAS to cope with autonomy, different beliefs, interests and desires of the agents that cohabit in the system. Based on the normative descriptions and the actual (past and present) actions of the agents, the system should detect the deviating behavior of any agent in the system.

According to Marin and Sartor(1999), there are four clear properties that can be ascribed to a norm of conduct. These are: applicability, (pre-)condition, effect and validity. The effect of a norm is what an agent must do: the obligation to carry out an action or a prohibition from carrying out an action. The condition is a situation that arises warranting the effect of the norm. The applicability of a norm is the time that a condition must hold in order for the effect to hold, while the validity of a norm is determined by its inclusion in the current body of code of conduct at that time.

We classify most of the norms of conduct that have appeared in the literature into two classes. *Low level* norms are those that specify a normative position or effect of a norm and the time when (or by which) an agent is expected to conform to it. Examples of works that formalize only low level norms include Sadri(2006), Stratulat *et al*(2001) and King et al(2017). We call this kind of norm representation as such because they instruct the agent on what to do and when to do it, rather than have the agent decide on when to act and for what reason. As it were, the condition for a norm's effect to take place is defined by a clock alarm signaling the arrival of the specified time. The violation of such a norm takes place when that time interval expires without the agent effecting the appropriate reaction stipulated by the norm's effect (i.e. either carry out a specific action or desist from doing so).

On the other hand there are *high-level* norms that specify an explicit description of both the effect of the norm and the condition warranting it as well as the relationship between the time of the effect and the time of the warranting condition. Examples of work from the literature that formalized only high level norms include the work of Governatori, Rotolo and Sartor(2005) as well as Artikis and Sergot(2010).

Our argument in this paper is that high-level norms are more appropriate for modeling real life norms of conduct than are low level norms. This is because low-level formalisms cannot capture many real life norms that require a detailed description of a warranting situation than simply the striking of the clock. In addition to this, real life norms of conduct also carry time constraint between the time the agents should effect his obligation or prohibition and the time of the warranting condition.

The violation of a high level norm is the failure of an agent to carry out the effect of the norm within the time constraint of the warranting situation. Furthermore, it is pertinent to determine the validity of the violated norm(s) as at the time the violation took place, because an invalid norm cannot be violated. A norm is said to be valid at a point in time if it is part (member) of the system of norms in consideration at that point in time. The interval, during which the norm is valid, is called the *external time* of that norm (Marin and Sartor, 1999).

A particular instance of low level norm representation is the formalism of King *et al.* (2017) which treats a normative fluent as a relationship between an event that must hold by a certain deadline, so that an obligation for an action a to be taken by a deadline d is the fluent *obl(a, d)*. Such an obligation is "discharged" if such an action takes place before the deadline or "violated" otherwise.

This paper presents a logic programming based knowledge representation formalism for norms that captures a norm's condition as a situation, its effect as an the obligation or prohibition of an action and the temporal constraint between them as a reified entity that captures qualitative and quantitative constraint. Section 2 discuses related works from the literature. Section 3 discusses the language of the formalism. Particularly, section 3.2.3 discusses the temporal constraint structures used to represent the reified temporal constraints expected between an agent's action and situation. Section 3.4 discusses how these temporal constraints can be checked using constraint satisfaction rules expressed as Logic programs. Section 3.5 shows how to use all these to make inferences about norm violation. Section 3.6 compares our formalism with some existing ones and points out parallel concepts from other papers.

## 2. NORM REPRESENTATION IN THE LITERATURE

According to Sergot(1990), there are two kinds of norms in legal reasoning. These are *qualification norms* and *norms of conduct*. Qualification norms define normative or legal properties and relationships of agents, while norms of conduct define how an ideal agent is expected to behave under certain circumstances. Consequently, unlike qualification norms, norms of conduct (or prescriptive norms) take cognizance of the fact that there is a difference between how an agent is expected to behave and how it might choose to behave in the real world. The departure between how an agent is expected to behave and how it eventually behaves is a *violation* of that expectation or norm.

Much work has been carried out on the logical representation of such norms of conduct. Notable papers in this regard include the works of Stratulat *et al* (2001), Marin and Sator(1999), Sadri *et al* (2006) and Panagiotidis *et al* (2009). As Artikis(2003) observed on the work of Marin and Sator, the paper focused on three major issues namely: norm applicability, norm effect and norm validity. Norm applicability describes the time interval during which a norm's (pre-) condition must hold in order for its effect to hold. The effect of a norm is the action an agent is expected to take or avoid taking when the norm's precondition takes place. Norm validity at a time describes the recognition of that norm as part of the body of legislation at a time. Marin and Sartor went on to identify the limitations of Kowalski and Sergot' s Event Calculus in solving these problems in the process of formalizing norms and presented an extension of Event Calculus for handling these problems.

In Stratulat et al (2001) the norms are represented by what seem like a reified ternary predicate. For example an obligation is a triple O(Agent, Action-type, Time-Interval) that carries the information that an agent is obliged to carry out an action of the type Action-type during the time interval. For example in Stratulat's language the statement

$$\text{holds}(j, O(ag, a, k)) \ldots\ldots\ldots \text{Axiom ST1}$$

states that an obligation *for an agent to carry out an action a at the time interval k* holds during time interval interval j. The interval k is the internal time of the norm in the language of Marin and Sator while the interval j is the external time during which the norm is in effect. Presumably k should be a subinterval of j. A violation of such a norm is inferred when the agent is found at a future time to have failed to carry out such an expected action within the prescribed time interval.

Similarly, Sadri et al(2006)'s representation of an obligation is the atom:

>    obliged(act(Act, Actor, Parameters), T, TC) or
>    prohibited(act(Act, Actor, Parameters), T, TC)

Thus an assertion that the actor *Actor* carries out (or is prohibited from carrying out) the action *Act* at time *T* which must satisfy the temporal constraint *TC*. Sadri et al's temporal constraints show the time points that bound by the time e g $10 > T > 17$.

In addition, for most of the norms represented by Sadri *et al*(2006), the norm's condition for an agent to be obliged to act or refrain from acting is set by time itself. Stratulat et al(2001)'s paper also shared this limitation. For example: *you can't park your car here between 10am and 5 pm*. Rather this kind of representation is inadequate for representing a norm like: *you can't park your car here from the moment it starts raining until it's 15 minutes after the rain*. While norms of the earlier type can be regarded as *low level norms* in the sense that they can only be applied within a specific time frame, those of the latter kind can be regarded as *high level norms*, in the sense that they apply anytime it rains. In that case it's up to the agent to decide whether the norm is applicable in a particular situation and at what particular time an agent finds itself in a deontic position.

A particular kind of low level norm is the normative fluent used by King et al(2017). Their normative fluent is written as *ob(a, d)*, denoting an obligation to carry out an action of type a before a deadline d. The problem with this kind of deadline norms is that such norms can be satisfied by actions of similar type undertaken before this norm becomes active. The problem with this representation is that unless the record of an action that discharged a similar fluent in the past is removed from the knowledge base as soon as it is discharged, it can inadvertently discharge future obligations.

This first gap is bridged in this paper by representing the time interval constraints using time point images (TPI), so that indefinite times and constraints can be represented. The second gap is also bridged here by using situations to model a norm's condition. A situation can be described by the propositional fluents (e.g. a raining fluent) that characterize it. Similarly a situation can be associated with a certain time in which it holds as done by Pinto (1994). This will make it possible to describe the conditions under which an agent cannot park his/her car for each of the two norm examples given above.

There are two major examples of norm representation formalisms that are indeed high level norms in the literature. The first is the temporalized formalization of legal norms by Governatori, Rotolo and Sartor(2005). In this formalization, a norm's condition has an implication relationship with the norm's effect (for example, the fact that an action is obligatory), with each side of the implication associated with relative times. A format for such rules looks like the following:

$$\text{condition} : t \Rightarrow \text{effect} : t \pm k$$

This rule reads like: *if the condition holds at time t, then the effect (such as an action becoming obligatory) holds at time t ± k*. The first problem with this kind of formalization is that it leaves no room for reasoning about violations of the norm (as pointed out by the authors themselves), particularly because there is no way of identifying the norm being violated. The best that can be done is to identify the violation of the effect of the norm such as an obligation to carry out an action at some time relative to the time the condition arose. The second problem with it is that temporal relations between conditions and effects can be more complex than can be represented by constraints involving two time points. Firstly when conditions arise, they usually have a starting point and a termination point. Therefore an agent may have an obligation to act at some point relative to (i.e. *before* or *after*) the start or finish or the time interval the condition lasted for. For example a robo-sweeper may be *obliged to start sweeping no later than 5 minutes after snowing stops*. Another example is when an agent is obliged to carry out an action *10 minutes before* the scheduled end of an event.

Another example of a formalism we consider appropriate for high level norms in the literature is one based on the Event Calculus and implemented in PROLOG by Aritkis and Sergot(2010). An example of a norm that can be represented in Artikis and Sergot's formalism is: *A faculty is empowered to supervise three years after earning a doctorate*. It is rendered thus:

$$\textit{holdsAt}(\textit{pow}(F, \textit{supervise}(F, X)) = \text{true}, T) \leftarrow$$
$$\textit{happens}(\textit{earnedDoctorate}(F) = \text{true}, T - 1096),$$
$$\textit{holdsAt}(\textit{enrolled}(X) = \text{true}, T).$$

The representation of the robosweeper example (i.e. *Robosweeper is obliged to start sweeping no later than 5 minutes after snowing stops*) however, reveals some problems with EC representations such as this one. It can be rendered thus in Artikis and Sergot's EC representation thus:

$$\textit{holdsAt}(\textit{obl}(\text{Robosweeper}, \textit{startweep}) = \text{true}, T) \leftarrow$$
$$\textit{holdsAt}(\textit{stop}(\text{snowing}), T1),$$
$$T1 \leq T,$$
$$T1 + 5 > T.$$

The problem with EC representations of norms stems from the fact that actions in EC are instantaneous. As such, when confronted with actions that have a duration such as sweeping, one can only capture the commencement and/or termination of such actions when necessary. A good representation of this norm however, should not only capture the commencement of the sweeping action, but the action in its entirety. The only way we can capture the sweeping action in an EC representation is to use the *start* and *stop* functions and treat sweep as an event like snowing. However, from an agent's point of view it is critical to distinguish between an event like *snowing* and an action like *sweeping*.

Nonetheless, Both Governatori et al as well as Artikis and Sergot(2010) in their representations captured the essence of a high-level norm which are a norm's condition and the essence of a temporal relationship between them. However, having a norm's condition and effect appear on either sides of an implication as done by both Governatori et al(2005) and Artikis and Sergot(2010), makes it impossible to give an identity to a particular high level norm.

In general many real life prescriptive norms are high-level norms. Besides, truly autonomous agents must be able to work with high level norms. Take for instance a real-life scenario. In an electronic institution where there is a norm that obliges an agent to supply an order within forty eight (48) hours after receiving the acknowledgement of payment for the order. The condition required for the agent to *carry out the supply of the order* (norm's effect) is *receiving the acknowledgment of payment for the order* (norm's condition).

Although this kind of norm can easily be formalized in the logic of Governatori et al, as:

$$\text{Pay-ack}(o) : t \Rightarrow \text{Obl}(\text{supply}(o)) : t + 48$$

We can only talk about the violation of the specific instances of the effect of this norm which is an obligation to supply an order at some given time (a low-level norm), and not the violation of the norm itself. Inferring which high level norm has been violated is critical to helping a sincere agent improve on its ability to be law-abiding. Communicating such norm violations also requires that each norm be identified so that the identity of the norm can be communicated to the erring agent.

Finally, inferring a norm violation requires that one determines whether or not the norm in question is active or not as only an active norm can be violated. This requires that we be able to have assertions about a norm's enactment and (if applicable) the repeal dates. This is only possible if each norm has an identity that can be used to refer to it. Such identities can even be quantified over in order to form general rules about norms.

Thus the representation of the norm used in this paper carries the following information:

> The identity of the agent involved,
>
> A normative fluent such as *obl(a)* or *pro(a)* that indicates that an obligation to carry out an action *a* or a prohibition from carrying it out. This is the norm's effect.
>
> A situation whose description will capture the nature of the condition warranting the effect of the norm,

The reified time constraint between the action and situation and
The norm's identity.

A normative token is treated as the relation bringing the other five elements together. This formalization eliminates the need for a normative implication of the sort championed by Governatori et al.

## 3. REPRESENTATION OF HIGH LEVEL NORMS

A norm is basically a rule. Each of these *normative rules* help in making inferences about what an agent is expected to do or desist from doing within a certain time frame. The inferences made from these normative rules are referred to as *normative tokens*.

A normative token is when an agent finds it is expected to effect a normative fluent, within some time constraint, and because of, the arising of some named situation in conforming to a named normative rule. A normative token relation is represented by a predicate NormPos thus:

NormPos : *Agent* × *NormFluent* × *Situation* × *TC* × *Norm-Id* → Boolean

(where the normative fluent is derived from an application of the function *obl* to an action type)

Norm-id is an identifier for the norm rule that produced the actual normative token which is an obligation. Every obligation that is produced by the norm bears the same norm-id. As such every rule that helps to infer obligations, prohibitions and permissions has a unique norm-id that it carries. This kind of rule naming is referred to by the term *rule reification* which is similar in spirit to the notion of Davidson's reification (Galton 1991), This is illustrated with Norm 3.1 and 3.2 as examples of rules generating this kind of normative tokens.

In sections 3.1 – 3.3, the language employed is a many sorted standard first order predicate logic with equality and with standard semantics for operators or ($\vee$), and ($\wedge$), negation ($\neg$) existential quantifier ($\exists$), universal quantifier ($\forall$) implication($\supset$) and equivalence($\equiv$). This is needed to express the properties of relations such as "partially describe" or "fully describe". In section 3.4 and 3.5, we resort to expressing norms with a language that is close to standard logic programs with equality and typical non-classical implication represented by *if* as well as, non-classical negation represented by $\neg$ .

Time is viewed as a linear and discrete model in which we recognize both time intervals and instants to be known as the sort Interval or Instants. The relations among two time points are represented by the infix predicates: $<, >, \leq, \geq$ with the signatures:

$<, >, \leq, \geq$ : Instant × Instant → Boolean

The beginning and ending of Intervals represented by the functions *begin* and *end* respectively; of Intervals are Instants as the signature below shows:

    *begin* : Interval $\rightarrow$ Instants
    *end* : Interval $\rightarrow$ Instants

In addition, we will use all of Allen(1984)'s binary interval relations represented by predicates: Meets, Before, Starts, Ends, During, Overlaps and their inverse relations. In addition, we will use other binary relations built on these such as Within (interpreted as Starts or During or Ends) and (improper) Subinterval (interpreted Within or Equals) by Koomen(1989).

**3.1 Normative Fluents and Action Types and Tokens**

The basic sorts introduced in this subsection are normative fluents and Action types as well as Action Tokens. A normative fluent is either the obligation on the part of an agent to carry out an action of a certain type or the prohibition on the part of an agent from carrying out an action of a certain type. The permission on the part of an agent to carry out an agent can be taken as the negation of the prohibition to carry out the same action. Because normative fluents are reified, the notion of obligation and prohibitions will be treated as the functions *obl* and *pro* respectively. These are both reckoned to have the following signature:

    *obl*, *pro*: ActionType $\rightarrow$ NormativeFluent

In other words, the *obl* and *pro* functions each take an action type as argument and returns a normative fluent. Apart from these there is a notion of negation for normative fluents which is the function ~ with the signature:

    ~ : NormativeFluent $\rightarrow$ NormativeFluent

The meaning of ~*obl*(a), where a is an action type a should be read as "permission not to carry out an action of type a", while the meaning of ~*pro*(a) should be taken to be "permission to carry out an action of type a". Norms that carry this kind of normative fluents cannot be violated.

Action types themselves carry a functional structure, so that *obl* and *pro* can be regarded as higher order functions. An action type is usually a function denoted by a verb that describes the generic type of the action, while its argument will be the object of the action. For example an action type *repair the car12* is the functional token *repair*(Car12). Thus *the obligation to repair the car12* is the normative fluent: *obl*(*repair*(Car12)).

So far we have been talking about an action type. However, there are instances when a norm's condition needs to be described by the occurring of an instance of an event type. In that case we define the time when an actual event takes place using a function *timeA* which given an action returns a time interval, with the signature:

$$timeA : Action \rightarrow Interval$$

Finally, assertions that make an actual event belong to a particular event type, are represented by the predicate Type-of-Action with the signature:

$$Type\text{-}of\text{-}Action : Action \times ActionType \rightarrow Boolean$$

### 3.2 Situations, Events, Processes

The basic sorts introduced here are Fluents, Situations, Events Processes and Time Intervals. In this paper, the conditions of a norm are modeled by situations. Our situation are similar to that described by Schubert(2000). A situation represents a complete state of the world with a time frame. Each *fluent* describes a specific aspect of a situation. In Schubert's FOL** language there are two kinds of relations that may exist between a fluent f and a situation s. In one relation a fluent completely describes the situation written as f **s, while in the other the fluent partially describes the situation f *s. In FOL**, the relation between a situation and a fluent that holds in it, is in general, an entailment relation i.e. s ⊨ f. Unlike FOL** however, our fluents which we will treat as functional tokens in our language are atomic propositions that do not require the use of quantifiers eg. Car17 is red, John loves Mary, etc. Speaking of functions, there is a time function for situations in this formalism which given a situation returns the time interval over which the situation persists. This function is *timeS* with the following signature:

$$timeS: Situation \rightarrow Interval.$$

In the reified situation calculus of Pinto(1994), the Holds predicate is used to represent the relation between a fluent and the situation in which it holds. Holds in Pinto's reified situation calculus does not in any way, suggest that a fluent completely describes a situation. As such we will thus use Holds(f, s) to connote the relation f *partially describes* s.

Schubert(2000) has argued the need for having situations that are completely described by specific fluents so that accurate causation relations can be expression between situations. The problem with only allowing situations to be partially described is that often causation relations are between certain aspects of situations. Other aspects of those situations are dormant in that relationship. This underscores the need for a *completely describes* relation between fluents and situation. Schubert renders this relation in FOL** as **. In order to respect the tradition from which this relation came from, we will use Holds** as relation between a fluent and the situation it completely describes so that *Holds**(f, s)* means f *completely describes* s. There are two specific relations between Holds and Holds** described below:

**Axiom 3.2.1**
*The relation "completely describes" is a specialization of "partially describes".*

$$\forall f, s.\ \text{Holds**}(f, s) \supset \text{Holds}(f, s)$$

**Axiom 3.2.2**
*If f partially describes s, there exists another situation of the same time as s, which f completely describes.*

$$\forall f, s.\ \text{Holds}(f, s) \supset \exists s_1.\ \text{Holds**}(f, s_1) \wedge timeS(s) = timeS(s_1)$$

We will define an implication relationship among fluents as the predicate ImplyF thus:

**Axiom 3.2.3**
*An implication relation exists among between a fluent and another, if the fact that it holds in any situation implies that the other holds in the same situation.*

$$\forall f, f_1.\ \text{ImplyF}(f, f_1) \equiv \forall s.\ \text{Holds}(f, s) \supset \text{Holds}(f_1, s)$$

A simple instance of the implication relationship between fluents will exist between a raining fluent and a wet fluent. The following axiom further clarifies the difference between the relations denoted by predicates Holds** and Holds.

**Axiom 3.2.4**
*If f completely describes a situation, then no other fluent other than those implied by f can hold in that situation.*

$$\forall f, s.\ \text{Holds**}(f, s) \supset (\forall f_1.\ f \neq f_1 \supset (\text{ImplyF}(f, f_1) \equiv \text{Holds}(f_1, s)))$$

Two special instances of fluents are the *occurring of an event* or the *progression of a process*. The fluents are represented by the functional structures: *occurring*(e) and *prog*(p) respectively, where *occurring* and *prog* are functions are applied on event token e and process p respectively giving rise to fluents with signatures:

    *occurring*: Event $\rightarrow$ Fluent
    *prog* : Process $\rightarrow$ Fluent

Each of these can partially or fully describe a situation. Just like situations both events and processes, have are eventualities with have an inherent time property. These are represented by time functions *timeE and timeP* for events and processes respectively, with the following signatures:

   *timeE* : Event → Interval
   *timeP* : Process → Interval

When an event or a process describes a situation, the time of that situation is the same as that of the event or process.

**Axiom 3.2.5**

*When the occurring of an event or the progression of a process describes a situation, the time of the situation is the same as the time of the event or that of the process as the case may be.*

  (a) Holds(*occurring*(e), s) ≡ *timeE*(e) = *timeS*(s)
  (b) Holds(*prog*(p), s) ≡ *timeP*(p) = *timeS*(s)

It is particularly important to be able to have situations completely described by these kinds of fluents, because such fluents actually give rise to the presence of other fluents. When the fact of an event occurring is a fluent that describes a situation, then the fact that certain agents play a role in that event is also a fluent that partially describes the situation. For the purpose of describing the relations between an actual event and an event type, as is needed in domain Axioms 3.2.6 and 3.2.7 below, we resort to using a predicate EventType thus:

**Axiom 3.2.6**
*If a classroom event occurs within a situation, then there must be a particular agent that plays the role of a teacher in that situation.*

  ∀e, s. Holds**(*occurring*(e), s) ∧ EventType(e, Class) ⊃
    ∃!a. Holds(*role*(a, Teacher), s)

**Axiom 3.2.7**
*If a classroom event occurs within a situation, then there must be at least one agent that plays the role of a student in that situation.*

  ∀e, s. Holds**(*occurring*(e), s) ∧ EventType(e, Class) ⊃
   ∃a. Holds(*role*(a, Student), s)

Finally it is important to distinguish between a process and events has been done in the literature and summarized in Galton(2016). Our definition of processes is close to that of Sowa(2000) as cited by Galton. A process can be discrete or continuous. Discrete processes each involves a finite sequence of either event occurrences or states (described by fluents). Each step can either be an event (or action), a state of affairs (like in a natural process like a fermentation or decay) or another process which is a process by itself. A continuous process can be treated as constituted by a single event. Herein lies the difference between our view of processes and Sowa's who breaks each continuous process into *initiation*, *continuation* and *termination* phases.

**Definition 3.2.8**
*A situation s is fully characterized by a process if Either that situation is fully characterized by either the occurring of an event Or there are two smaller situations whose times cover s and the first situation is fully characterized either by the occurrence of an event or a fluent and the second is also a process.*

$$\forall p, s.\ Holds^{**}(prog(p), s) \equiv$$
$$(\exists e\ Holds^{**}(occurring(e), s)) \vee$$
$$\exists s_1.(\exists e\ Holds^{**}(occurring(e), s_1) \vee \exists f\ Holds^{**}(f, s_1)) \wedge$$
$$\exists s_2, p_1.\ Holds(prog(p_1), s_2) \wedge Cover(timeS(s_1), timeS(s_2), timeS(s))$$

(where $\forall j, k, m.\ Cover(j, k, m) \equiv Starts(j, m) \wedge Finishes(k, m) \wedge Meets(j, k)$
see (Koomen 1989) )

Either way, Axiom 3.2.7 above captures the definition of a process.

### 3.3 Temporal Constraints

The basic sorts introduced here are time point images (TPI) and Time Constraints (TC). A TPI is a relative point with an anticipated time interval. It can be the beginning (B), End (E) or any relative time point within the anticipated time interval. Therefore a TPI may be either of the constants {B, E} or an application of the time displacement function *tdisp* to either B or E and an integer. B and E signify the beginning and end of a time integer. The signatures are given below:

$tdisp$: TPI $\times$ Integer $\rightarrow$ TPI

Our approach to composing constraints here is therefore is to let each basic constraint relate some time point image (i.e. its beginning, end, or some point in between) of the time interval of the effect of a norm which is the time of the action of the type specified, with some time point image of the effect of the norm. A basic temporal constraint is composed by the application one of the functions such as *eq* (equal), *ge* (greater than or equal to), *le* (less than or equal to), *gt* greater than) and *lt* (less than) to an ordered pair of TPI, the first relating to the action (or norm's effect) while the second is related to the situation (norm's condition) with the signatures:

$eq, ge, gt, le, lt$ : TPI $\times$ TPI $\rightarrow$ TC

The functions *and*, *or* and *not* introduced here are used to make temporal constraints with natural Boolean meanings and the following signatures:

$$not : \text{TC} \to \text{TC}$$
$$and, or : \text{TC} \times \text{TC} \to \text{TC}$$

For example:

*eq*(B, E) means the first interval ends exactly when the second interval begins.

*eq*(*tdisp*(B, 3), E) means the second interval ends 3 units into the beginning of the first interval.

Other temporal constraints can be composed from other temporal constraints by the functions and, or as well as neg. For example, *and*(*eq*(B, B), *eq*(E, E)) describes two intervals that are equal.

Each of Allen(1984)'s qualitative relations can be represented by our TCS. The following are the equivalences between Allen's interval relations and our TCS.

| | |
|---|---|
| *Before* is equivalent to | *lt*(E, B) |
| *Overlaps* is equivalent to | *and*(*lt*(B, B), *lt*(E,E)) |
| *Contains* is equivalent to | *and*(*lt*(B, B), *gt*(E, E)) |
| *Starts* is equivalent to | *and*(*eq*(B, B), *lt*(E, E)) |
| *Finishes* is equivalent to | *and*(*gt*(B, B), *eq*(E, E)) |
| *Meets* is equivalent to | *eq*(E, B) |

A major advantage of the TCS representation is that it can represent constraints that combine both qualitative and quantitative relationships. Examples of such relationships are:

*Starts not later than 4 units of time into* is equivalent to *le*(B, *tdisp*(B, 4))
*Ends 10 units of time into* is equivalent to *eq*(E, *tdisp*(B, 10))

These are the kinds of constraints that many real-life norms may contain as holding between norm conditions and their effects.

Finally we must state that in using the context of normative positions, each basic constraint's argument is pair of the action-type's time interval and the situation's time interval. For example the following norm, identified by Norm001, means the agent Robosweeper has an obligation to begin sweeping snow not later than 4 time units after a situation in which it snowed.

NormPos(Robosweeper, *obl*(Sweep_Snow), Snowing, *and*(*gt*(B, E), *le*(B, disp(E,4))), Norm001)

**3.4 Normative Rules and Rule Reification**

In our formalism, norms or normative rules are represented as Logic programs in which the heads are normative tokens and the bodies contain the description of norm's conditions which are represented by situations. i.e.

>Normative-Token (NT) if
>>Description of the situation that appears in NT

Each normative token is an assertion that an agent is to conform to a normative fluent, in some named situation (which is later described in the body of the clause), within some temporal constraint.

This representation eliminates the need for a normative implication in the sense used by Governatori et al(2005) and Artikis and Sergot(2010) because unlike these previous formalisms in which a norm's condition and its effect are antecedent and consequent in an implication. This is achieved by making the norm's condition, some situation s and the norm's effect, some normative fluent, terms for the 5-tuple normative token(reification). Thus when a situation arises whose description matches what appears in the body of the normative rule, a normative token is inferred which is derived from the head of the normative rule.

This reification of both the condition and effect of norms is the main difference between this formalism and that of Governatori et al. and Artikis and Sergot. For these both of these formalisms, it is virtually impossible to reify a norm's conditions without resorting to situations as we have done.

The last term of a normative token is an identifier for the normative rule that produced it. This value assigned to this term is the identification for the normative rule from which the normative token was inferred. Although the identifier only appears in the head of the normative rule, it is in fact an attribute of the normative rule and it is the identifier for the rule. Indeed every normative token derived from the same normative rule bears that same identity. This is a special kind of reification for rules we shall call *rule reification*.

As we will see from the examples in the next section, it is often necessary to describe other situations that have some qualitative relations with the situation being described.

**3.5 Real life Norms**

In this section we present our formalism's representation of some real life norms. It is a rather trivial exercise to show that the formalism is able to represent all low-level norms used to test a formalism such as that Stratulat et al (2001). We will return to that question later. At this point our focus is on representing high level norms.

**Norm 3.1**
*A teacher assigned to teach a class must arrive either on time or not later than 10 minutes*

*into the time of a class.*

∀a, v, s,
NormPos(a, *obl(arrive-at(v))*, s, *and(le*(E, *tplus*(B,10)), *ge*(E, B)), OB101) if
    ∃ e.
        Holds**(*occurring*(e), s) ∧
        EventType(e, Class) ∧
        Holds(venue(e, v), s) ∧
        Holds(role(a, Teach), s)).

**Norm 3.2**
*Student must register for his/her courses in a semester within one month of the commencement of the semester.*

∀a, o, s.
NormPos(a, *obl(register-for*(a,sem))), s, *and(and(ge*(E, B), *le*(E,B, 30)), *ge*(B,B)), OB102) if
    ∃$s_1$, $s_2$.Holds**(*studentship*(a),$s_1$) ∧
        ProcessType(sem, Semester) ∧
        Holds**(*prog*(sem), s) ∧
        Within(*timeS*(s), *timeS*($s_1$))∧
        ¬ Holds(on-suspension(a, sem), s).

A prohibition is a norm that disallows an agent from carrying out an action of a certain type within some time interval that has a temporal relation with the warranting situation (condition) of the prohibition. The structure of a prohibition is the same as that of a normative position except that the normative fluent is derived from the application of the *pro* function to an action type.

The norm 3.3 and norm 3.4 illustrate examples of rules that infer such normative tokens.

**Norm 3.3**
*Student must not be allowed to come in for an examination thirty (30) minutes after the commencement of the examination.*

∀a, s, e.
NormPos(a, *pro(arrive*(v)), s, *gt*(E, *tplus*(30)}, PR0103), if
        Holds**(*occuring*(e), s) ∧
        EventType(e, Examination)) ∧
        Holds(*venue*(e,v), s) ∧
        Holds(*role*(a, Candidate), s).

**Norm 3.4**
*It is prohibited for members of university community to release confidential document of the university to public domain without authorization.*

∀a, doc, s,
*Prohibition*(a, *pro(release*(doc)), s, *or*(*and*(*lt*(B, E), *ge*(B, B)), *and*(*le*(E, E), *gt*(E, B))), PRO014) if
    ∃$s_1, s_2$, u.
    Holds**(*alive*(doc), $s_1$) ∧
    Holds**(*statusdoc*(doc, Confidential),s) ∧
    Holds**(*employ*(u, a), $s_2$) ∧
    University(u) ∧
    Owns(doc, u) ∧
    Subinterval(*timeS*(s), *timeS*($s_1$)) ∧
    Subinterval(*timeS*(s), *timeS*($s_2$)) ∧

A permission is a norm that allows an action by an agent as a result of a certain situation arising. In this formalization, we will treat permissions as the fluent negation of prohibitions denoted by the function ∼.

Examples of this is given in norm 3.5 and norm 3.6.

**Norm 3.5**
*Members of the university community are permitted to put on their official identity card while on duty.*
∀a, s.
NormPos(a, ∼*pro*( *put_id-on*(a)), s, *and*(*ge*(B, B), *le*(E, E)), PER010) if
    ∃u, $s_1$
    Holds**(*onDuty*(a), s) ∧
    Holds**(*employ*(a, u) $s_1$) ∧
    University(u) ∧
    Within(*timeS*(s), *timeS*($s_1$)

**Norm 3.6**
*Lecturer is permitted to give reading books on his assigned course to the student at the beginning of lecture in a semester.*
∀a, b, s.
NormPos(a, ∼*pro*(*give*(b)), s, *le*(E, B), PER012) if
    ∃co.
        Holds**(*prog*(co), s) ∧
        ProcessType(co, Course) ∧
        Holds(*role*(a, Teacher), s) ∧
        Holds(*role*(b, ReadingBook), s)

Having represented norms, it is important to discuss the representation of the validity of those norms. Norm validity is determined by the date of enactment and . Let us take as an example the constitutive norm from the formalization of the British Nationality Act 1981 (Segot *et al*., 1986).

The norm *any person born in the UK becomes a British citizen* is a norm that was only valid until 1981. A representation of that norm in the proposed language is presented thus:

$\forall$x, s
NormPos(HMG, *obl*(*grant-citizenship*(x)), s, *eq*(E, B), NBB-1) if
     Holds**(born-in(x, UK), s)

However, any agent interpreting that norm on behalf of Her Majesty's Government (HMG) must be aware that the validity of the norm is from 1950 and 1981 i.e.

    Enact(NBB-l, 1950)
    Repeal(NBB-1, 1981)

As such it is only if the norm situation s happened within that interval of validity that the norm NBB-1 is valid. Validity is an important condition for deciding norm violation as we shall demonstrate.

### 3.5 Satisfying Temporal Constraints

Our attention now turns to determining when two time intervals satisfy a certain temporal constraints and for this purpose, we will define constraint satisfaction rules (CSR) that will determine whether an ordered pair of intervals satisfies a temporal constraint.

Intuitively, it is easy to see how the satisfaction of any of our basic constraints can translate into logical relations that involve two time points. Similarly the satisfaction of a compound constraint can be broken down into the satisfaction of constituent constraints.

CSRs for compound temporal constraints are recursive in the sense that they are composed from other CSRs as illustrated by the following axiom:

**Axiom 3.5.1**
$\forall tc_1, tc_2, j, k.$
Satisfy-Cons( j, k, *and*($tc_1$, $tc_2$)) if
       Satisfy-Cons( j, k, $tc_1$) $\wedge$
       Satisfy-Cons(j, k, $tc_2$).

Two similar rules exist for the *or* function. That is embodied in axiom 5.2(a) and (b) below:

**Axiom 3.5.2** (a)
$\forall j, k, tc_1, tc_2.$
    Satisfy-Cons( j, k, *or*($tc_1$, $tc_2$)) if
       Satisfy-Cons(j, k, $tc_1$)
**Axiom 3.5.2** (b)
$\forall j, k, tc_1, tc_2.$

Satisfy-Cons(j, k, *or*(tc$_1$, tc$_2$))  if
  Satisfy-Cons( j, k, tc$_2$)

The other non-recursive rules handle basic temporal constraints. These rules represent a simple translation of each of the temporal constraint functions into the equivalent relation. The following axioms 5.3 to 5.6 are all examples of the *lt* constraint.

**Axiom 3.5.3**
$\forall$j, k, t$_1$, t$_2$.
  Satisfy-Cons(j, k, *lt*(*tdisp*(B, t$_1$), *disp*(B, t$_2$)) ) if
    *begin*(j) + t$_1$ < *begin*(k) +t$_2$

**Axiom 3.5.4**
$\forall$ j, k.
  Satisfy-Cons(j, k, *lt*(B, E)) if
    *begin*( j) < *end*(k)

**Axiom 3.5.5**
$\forall$j, k.
  Satisfy-Cons(j, k, *lt*(E, B)) if
    *end*(j) < *begin*( k)

**Axiom 3.5.6**
$\forall$j, k.
  Satisfy-Cons(j, k, *lt*(E, E)) if
    *end*(j) < *end*(k)

The next section discusses norm how constraint satisfaction rules is used to infer norm violations.

## 3.6  NORM CONFORMANCE AND VIOLATIONS

In order to make inferences about norm violation we need to know the existence of three fundamental conditions. The first condition will be the existence of a normative token such as an agent's obligation or prohibition to carry out or desist from carrying out an action. The second condition will be the failure to conform to the requirement of the normative token. The third condition is the validity of the normative rule at the time the condition held.

The condition for a norm to be valid with respect to a situation is for the situation to take place while the norm's enactment remains in force. This is formalized as:

**Axiom 3.6.0**
$\forall$norm-id, s.
Validwrt(norm-id, s) if
  Enact-at(norm-id, t) $\wedge$
  t <= begin(*timeS*(s)) $\wedge$
  $\forall$t$_1$. $\neg$(t $\leq$ t$_1$ $\leq$ end(*timeS*(s)) $\wedge$ Repeal(norm-id, t$_1$))

The conformance of an agent to a norm requiring an agent's obligation requires the existence of an action of the expected type to be carried by the agent within the required time limit. This is formalized by the following clause:

**Axiom 3.6.1**

$\forall$a, norm-id, s.
Conform(a, norm-id, s) if
$\quad\quad\exists$ act-type, tc, act.
$\quad\quad$NormPos( a, *obl*( act-type), s, tc, norm-id) $\land$
$\quad\quad$Type-of-Action(act, act-type) $\land$
$\quad\quad$Actor(a, act) $\land$
$\quad\quad$Validwrt(norm-id, s) $\land$
$\quad\quad$Satisfy-Cons(*timeA*(act), *timeS*(s), tc).

The violation of an obligation takes place when in the occurrence of the situation within the validity period of the rule that generated the normative token, the agent implicated is unable to carry out the needed action within the required time constraint. That is formalized as the following clause:

**Axiom 3.6.2**

$\forall$a. norm-id, s.
$\quad$Violate(a, norm-id, s) if
$\quad\exists$ act-type, tc.
$\quad$NormPos(a, *obl*(act-type), s, tc, norm-id) $\land$
$\quad$Validwrt(norm-id, s) $\land$
$\quad\quad\quad\forall$act $\neg$ (Type-of-Action (act, a-type) $\land$
$\quad\quad\quad$Actor(a, act) $\land$
$\quad\quad\quad$Satisfy-Cons(*timeA*(act), *timeS*(s), tc) )

In the case of a prohibition, a violation takes place when in the case of the occurrence of the situation, the implicated agent carries out the forbidden action during a time interval that satisfies the constraint with the time of the situation. Such a violation is formalized thus:

**Axiom 3.6.3**

$\forall$p, norm-id, s.
Violate(a, norm-id, s) if
$\quad\quad\exists$ act-type, tc, act.
$\quad\quad$NormPos( p, *pro*(act-type), s, tc, norm-id) $\land$
$\quad\quad$Type-of-Action(act, act-type) $\land$
$\quad\quad$Actor(a, act) $\land$
$\quad\quad$Validwrt(norm-id, s) $\land$
$\quad\quad$Satisfy-Cons(*timeA*(a), *timeS*(s), tc).

Similarly conformance to a norm that is a prohibition is the absence of the prescribed action within the prescribed time constraint. That is contained in the following axiom:

**Axiom 3.6.4**
∀a. norm-id, s.
Conform(a, norm-id, s) if
    ∃ act-type, tc.
    NormPos(a, *pro*(act-type), s, tc, norm-id) ∧
    Validwrt(norm-id, s) ∧
    ∀act ¬(Type-of-Action (act, act-type) ∧
        Actor(a, act) ∧
        Satisfy-Cons(*timeA*(act), *timeS*(s), tc) )

However, an agent needs not do anything else to conform to a permission to either carry out or avoid an action. That idea is formalized in Axiom 3.6.5.

**Axiom 3.6.5**
∀a, norm-id, s.
Conform(a, norm-id, s) if
    ∃act-type, tc. NormPos(a, ~*pro*(act-type), s, tc, norm-id)

**Axiom 3.6.6**
∀a, norm-id, s.
Conform(a, norm-id, s) if
    ∃act-type, tc.
    NormPos(a, ~*obl*(act-type), s, tc, norm-id)

This approach contrasts with timed violations that was implemented by Stratulat et al[21]. In their work, a violation of an obligation is said to have taken place at a time t, if an agent has an obligation to carry out an action during the time interval ($t_1$, $t_2$) and as at time t which is later than $t_2$ the action has not yet been taken. We believe that this approach of theirs is flawed because a violation is best understood as occurring under some particular circumstances rather than as occurring at some particular time. For example it is more meaningful to infer that:

> Mr X violated the of promptness obligation with regard to the CSC777 class of 1 April 2016

than to infer that:

> Mr X is reckoned to have violated an obligation to report at the Faculty Lecture theatre between 10 and 10:10 on 1 April 2016.

4. **COMPARISON WITH EXISTING FORMALISMS**

The norm representation formalism that is closest to that developed in the paper is the one offered by Panagiotidi, Nieves and Vazquez-Salceda(2009). In their representation, a norm is a 6-ary relation which consists of:

The actor
The (deontic) modality, which may either be an *obligation* or a *permission*.
The activating condition for the norm
The maintenance condition for the norm
Deontic Statement and,
The deactivating condition

The activating condition is the equivalent to the norm's precondition. The deontic statement is a description of the state change that the agent under obligation is expected to effect. The deactivating condition is a sign that the state change has been effected. A norm becomes activated when the activating condition becomes true and remains activated until the deactivating condition becomes true. The deactivating condition is satisfied when the agent has successfully effected the requirement of the norm. The maintenance condition captures the deadline for an active norm to become deactivated. Therefore a violation occurs when an active norm remains active when its maintenance condition has become untrue (i.e. the deadline has been passed while the deactivating condition is yet unsatisfied by the agent).

However there are significant differences between the approach presented in this paper and their approach. Firstly, the representation in this paper does not only capture activating conditions; rather, it captures named instances of their occurrences as *situations*. So that it is possible to have named instances of an event or process type as a situation. That is not the case with the Panagiotidi *et al*'s representation. Again their inference of violation is based on the observing changes in condition and not necessarily on the occurrence or non-occurence of actions.

The norm in Panagiotidi et al about the obligation of a repair company rc to repair a car within 4 days of taking the car to the garage will be rendered thus:

NormPos(rc, *obl*(*repair*(car17)), s, *and(ge(B, B), le(E, disp(E, 4)))*, OBRC1) if
Holds(*arrive*(car17, Garage), s)

We wish to note that our representation allows us to state the car repair that will satisfy the new obligation cannot precede the arrival of the car at the garage. This way we can avoid mistaking a previous repair done on car17, with the repair required for a new visit to the garage. This is an advantage that our formalism has over formalisms such as King et al (2017) and Panagiotidi et al (2009) that depend on deadlines.

Hashmi, Governatori and Moe(2016) present a taxonomy of obligations that expands on a simple categorization of obligations due to Governatori (2010). A *maintenance* obligation is one in which a norm's effect must continue throughout the time of the obligation (or some specified situation). The norm represented by example 2 in this paper is one such norm.

It is also the case that 4.1 is an example of Governatori's *achievement* obligation.

An *achievement* obligation is one in which an actor is expected to carry out the action at least once within the time limit. An example of this which can be represented in the language of this paper is:

> *The Prime Minister must pay the Queen a visit during the parliamentary break.*

The representation for this in the representation language presented in this paper would treat the parliamentary break as an event whose occurring constitutes the warranting situation and the agent whose visitation action to the Queen must fall within that situation is the Prime Minister.

Finally in this section, in order to demonstrate that our formalism is versatile enough to represent low level norms, we present a representation for a norm from Stratulat et al (2001) that *obliges a taxpayer to pay up within the first 31 days of the year*, thus:

NormPos(a, *obl*(pay-tax), s, *or*(*eq*(E, B), *le*(E, *disp*(B, 31)), OBLTAX1) if
$$\exists y(\ begin(timeS(s)) = (January, 1\ y) \land$$
$$end(times(s)) = (December, 31, y)) \land$$
$$Holds(taxpayer(a), s).$$

The good side of our representation is that any violation can report on the year for which a violation took place and the identifier for the particular tax code violated, which in representation is OBLTAX1.

## 5. CONCLUSION

This paper has identified two broad categorizations for prescriptive norms in the literature. High level norm representations capture the norm's condition and effect as well as the temporal constraints between them. The norm's condition is treated as a situation warranting the need for an agent to either carry out or refrain from carrying out an action of a certain type. The effect of the norm is the action that the agent is obliged or permitted to take or prohibited from taking.

This contrasts with other approaches, which accounts a norm as the obligation or permission to take an action or the prohibition from taking an action within a particular time interval. The key disadvantage of low-level norms is that an agent's autonomy is curtailed if s/he needs an external prescription in order to know what actions s/he is expected or allowed to take at specific times. On the other hand, an agent with a high level norm representation of norms can determine from that representation, at what times an actions they have an obligation to act or prohibited from acting. The difference between the two approaches is somewhat like that between low-level programming and high-level programming.

The major disadvantage of high level norms as they appear in the existing literature is that generalized violation inference is practically impossible. The formalism for high-level norm representation presented in this paper fixes that problem by relaxing the strict normative implication relationship between a norm's condition and its effect as done by Governatori et al(2005) and Artikis and Sergot(2010). The formalism also makes it possible for norms to be identified through *rule reification*.

Another key advantage of low level norms it easily enables an agent to detect when there are contradictory norms as shown by Artikis(2003). A contradiction occurs when there two different norms, one of which obliges an agent to carry out an action of a certain type, and the other prohibits actions of the same type at around the same time. In order to detect contradictory norms, there is a need for a logical mechanism to convert high level norms into low level norms on an as needed basis.